\documentclass[12pt,fleqn]{article}
\usepackage{a4wide}
\usepackage{natbib}
\usepackage{amssymb, amsmath}
\usepackage{amsfonts}
\usepackage{amsthm}
\usepackage{graphics}
\usepackage{bm}
\usepackage{float}
\usepackage{appendix}
\usepackage{mathpazo}
\usepackage{comment}
\usepackage[multiple]{footmisc}
\usepackage{rotating}
\usepackage{subcaption}
\usepackage{booktabs}
\usepackage{threeparttable}
\usepackage{pdflscape}
\usepackage{enumerate}
\usepackage{setspace}
\usepackage[normalem]{ulem}
\usepackage{bm}
\usepackage{url}

\usepackage[margin=0.75in,paper=letterpaper]{geometry}

\newtheorem{theorem}{Theorem}

\theoremstyle{definition}
\newtheorem{assumption}{Assumption}[section]

\numberwithin{equation}{section}
\numberwithin{lemma}{section}
\numberwithin{corollary}{section}
\numberwithin{remark}{section}
\numberwithin{theorem}{section}
\numberwithin{proposition}{section}

\begin{document}
\title{{\sc Threshold Regression for Fixed-$T$ Panel Data with Interactive Fixed Effects}\thanks{The authors would like to thank Heino Bohn Nielsen (Editor), and two anonymous referees for valuable comments and suggestions. Thank you also to the Knut and Alice Wallenberg Foundation for financial support through a Wallenberg Academy Fellowship.}}
	
\author{Jan Ditzen\thanks{Email: jan.ditzen@unibz.it}\\
	{\small Free University of Bozen-Bolzanno} \and Yiannis Karavias\thanks{Corresponding author: Department of Economics, Finance and Accounting, Brunel University of London, Uxbridge, UB8 3PH, London, UK. Tel: +44 (0)1895 267545. E-mail address: \texttt{yiannis.karavias@brunel.ac.uk}.}\\
	{\small Brunel University of London}\\
	{\small and}\\
	{\small University of Birmingham} \and Joakim Westerlund\thanks{Email: joakim.westerlund@nek.lu.se}\\
	{\small Lund University}\\
	{\small and}\\
	{\small Deakin University}}
	
\maketitle

\begin{abstract}
	This paper develops a new toolbox for estimation and inference in panel data threshold regression models with interactive fixed effects and a fixed number of time periods, $T$. The toolbox is designed to be simple, accurate and computationally efficient. It is based on a simple least squares style estimator of the model parameters, and includes a number of inferential procedures for testing hypotheses regarding not only the threshold but also other parameters. The new toolbox is applied to study the impact of inflation on economic growth.
\end{abstract}


\smallskip 

\noindent \textbf{Published at the Oxford Bulletin of Economics and Statistics}: 

\url{https://doi.org/10.1111/obes.70102}
\doublespacing	
\section{Introduction}\label{sect:intro}

In their discourse on threshold models, Tong and Lim (1980) coined the term ``practical nonlinear models'' to describe models which are capable of capturing nonlinear relationships but whose interpretation is still intuitive and convenient to empirical researchers. Threshold regression models belong to this category and have found fertile ground in economics (and elsewhere) because the idea of different regimes and regime-separating thresholds fits well with many theories and empirical observations. For example, the macro economy is characterised by expansions and recessions, or ``good times'' and ``bad times'' as they are frequently called. Inflation targeting central banks, such as the Federal Reserve, the Bank of England and the European Central Bank will take action if inflation crosses certain pre-determined thresholds. Another example is the evidence that exists of threshold effects in public debt-to-GDP, which tilted the economic debate to pro-austerity policies in the aftermath of the global financial crisis. Reviews of the widespread use of threshold models in economics can be found in, for example, Hansen (2011), or Tong (2015).

A critical condition for threshold regression models to be estimable is that there is enough variation in the data to be able to separate one regime from another. This calls for the use of lengthy time series (or cross-section) data with sufficient observations on each regime, a condition that is rarely met in practice.\footnote{Some studies average their data over subperiods to eliminate short-term fluctuations and apply existing techniques for threshold regression to the resulting averaged data (see, for example, Khan and Senhadji, 2001, and Kremer et al., 2013). In these studies, the effective number of time series observations is typically very small.} Also, even if such lengthy time series were available, there is no guarantee that they are informative enough to ensure reliable estimation of the threshold (see, for example, Chudik et al., 2017). Panel data have the advantage that they allow one to pool the information contained in multiple time series, which means that accurate estimation is possible even if the number of time periods, $T$, is relatively small. This is important since while the number of time periods cannot be increased other than by the passage of time, statistical agencies keep publishing time series data for individuals, firms and countries. Thus, while $T$ is usually quite small, the number of cross-sectional units, $N$, can potentially be very large.

Hansen (1999) was among the first to make the above point. He considered a static panel data regression with slope coefficients that may be subject to threshold effects. The main finding is that asymptotically valid estimation and inference is possible even if $T$ is fixed, provided only that $N$ grows large. The required number of time periods is therefore greatly reduced when compared to the pure time series case, and this has in turn facilitated widespread use of threshold regressions. As an indication of this, out of the over $6,300$ citations that Hansen's study has attracted to date, a vast majority are applications to panels in which $T$ is small and only $N$ is large.

In the present paper we take Hansen's work as our starting point. Our point of departure is his preference to assume that the panel data is cross-sectionally independent, which is unrealistic. As is well-known, a major advantage of using panels as opposed to time series or cross-section data is the ability to deal with unobserved heterogeneity, which can be detrimental if correlated with the regressors. Hansen (1999) allows for cross-section unit-specific fixed effects, which are eliminated prior to the estimation of the model by transforming all variables into deviations from their time series means. This type of demeaning can be quite effective in reducing the degree of serial correlation but it does not help if there is correlation across the cross-sectional units. This observation recently motivated Barassi et al. (2023), and Miao et al. (2020) to extend Hansen's (1999) fixed effects specification to one in which the unit-specific fixed effects, called ``factor loadings'', are allowed to interact multiplicatively with common time-specific effects, called ``common factors''. The resulting ``interactive fixed effects'' specification is very flexible since the factors are allowed to affect each cross-sectional unit differently, and is in fact implied by many economic theories.\footnote{For a detailed introduction and a practitioner's guide to interactive fixed effects, see Ditzen and Karavias (2025).} However, in these studies $N$ and $T$ are both assumed to be large, which is again rarely the case in practice.

The present study is motivated by the above discussion. The purpose is to develop a new toolbox for panel threshold regression that is robust to the presence of interactive fixed effects, yet does not require $T$ to be large. This is a challenge because, as already alluded to, accurate estimation of complex models such as the one considered here standardly require large sample sizes. Another challenge is that with ever more complex models being estimated using ever larger data sets, computational aspects are of increasing importance, and increasingly receive attention, just not in economics. For instance, the maximum likelihood approach of Miao et al. (2020) requires grid search to estimate the threshold parameter where every point in the grid involves solving a high-dimensional, non-convex optimization problem to estimate the remaining parameters, and singular value thresholding to estimate the number of factors. As a result, the approach is not only computationally very costly but can also be difficult to get to converge, and even if it does converge it may not be to the global optimum.

Tackling the above challenges requires attention to detail and optimizing both performance and computational speed at each step. Estimating the threshold parameter necessitates searching over a grid of up to $NT$ points. It is therefore important that this search is done as efficiently as possible and that all other calculations are kept to a minimum. Hansen (1999, 2000) recommends using least squares (LS) to estimate the threshold parameter, and therefore so do we. However, instead of minimizing the LS residuals, which will generally lead to inconsistency because of the unattended interactive fixed effects, we minimize the residuals obtained by applying the common correlated effects (CCE) estimator of Pesaran (2006). The reason for focusing on CCE as opposed to any other interactive fixed effects estimator is that it has a simple closed form, it does not require accurate estimation of the number of factors, it performs very well in small samples, and it is valid even if $T$ is fixed (see Westerlund et al., 2019). We then use this estimator as a basis for constructing hypotheses tests and confidence intervals for all model parameters. As far as we are aware, the resulting toolbox is the first to enable asymptotically valid estimation and inference in fixed-$T$ panel threshold regression with interactive fixed effects. The community-contributed command \texttt{xtthreshold} implements the new toolbox in Stata.

The usefulness of the new toolbox is illustrated using as an example the relationship between inflation and economic growth. There is by now considerable empirical evidence to suggest that low inflation rates have a positive impact on growth which turns negative as inflation increases (see Azam and Khan, 2022, and Nell, 2023). However, most of this evidence ignores the fact that all countries have access to the same pool of technological knowledge. This knowledge can be seen as a common factor with loadings that measure the extent to which countries have access to it (see Pesaran, 2007). Our main concern here is that technology knowledge is likely correlated with inflation, as higher productivity is expected to lead to lower prices, in which case estimation by standard techniques will be misleading. The data that we use cover 74 countries from 1970 to 2022. According to the results, while beneficial below this threshold, depending on the countries being considered, inflation in excess of 1.1\%--3.5\% is harmful for economic growth.

The rest of the paper is organised as follows. Section \ref{sect:mod} introduces the model and the new toolbox, whose asymptotic and small-sample properties are investigated Sections \ref{sect:asy} and \ref{sect:mc}, respectively. Section \ref{sect:appl} contains the empirical application. Section \ref{sect:concl} concludes. All proofs are relegated to an online appendix.

\section{Model and toolbox}\label{sect:mod}

We consider the following panel threshold regression model:
\begin{align}
	y_{i,t} & = \bm{\beta}'\bm x_{i,t}+\bm \delta'\bm h_{i,t}(\gamma)+e_{i,t},\label{eq:y}
\end{align}
where $\bm x_{i,t}:= (x_{1,i,t},...,x_{k,i,t})'$ is a $k\times 1$ vector of regressors, $\bm h_{i,t}(\gamma)$ is a $l\times 1$ vector of regressors whose value depends on the threshold parameter $\gamma$, and $e_{i,t}$ is an error term. We assume that $\bm h_{i,t}(\gamma) :=\bm S'\bm x_{i,t} \mathbb{I}(z_{i,t}>\gamma)$, where $a:=b$ means that $a$ is defined by $b$, $\mathbb{I}(A)$ is the indicator function taking on the value one if the event $A$ is true and zero otherwise, $z_{i,t}$ is a scalar threshold variable, and $\bm S$ is a $k\times l$ matrix of rank $l$ that selects the elements of $\bm x_{i,t}$ whose coefficients are subject to change. As in the bulk of the previous literature (see, for example, Hansen, 1999, 2000), we assume that $\bm S$ is known and that the threshold variable $z_{i,t}$ is one of the regressors in $\bm x_{i,t}$.\footnote{If $z_{i,t}$ is not a part of $\bm x_{i,t}$, after a rearrangement of the order of the observations, the threshold model becomes equivalent to the structural breaks model of Karavias et al. (2023) and the same asymptotic theory for estimation and testing applies.} In our growth application, $z_{i,t}$ is inflation, which is also a regressor in $\bm x_{i,t}$.

A key feature of the above model is that the regressors and hence also the threshold variable are not required to be exogenous but are instead allowed to be correlated with the error $e_{i,t}$ via the following common factor structure:
\begin{align}
	e_{i,t}& =\bm \mu_{i}'\bm f_{t}+\varepsilon_{i,t}, \label{eq:e}\\
	\bm x_{i,t}& = \bm \Lambda_{i}' \bm f_{t} +\bm v_{i,t},
	\label{eq:x}
\end{align}
where $\bm f_{t}$ is a $m\times 1$ vector of common factors with $\bm \mu_{i}$ and $\bm \Lambda_i$ being $m\times 1$ and $m\times k$ matrices,  respectively, of factor loadings. The scalar $\varepsilon_{i,t}$ and $k\times 1$ vector $\bm v_{i,t}$ are idiosyncratic error terms that are independent of each other. The interactive fixed effects are here given by the products $\bm \mu_{i}'\bm f_{t}$ and $\bm \Lambda_{i}' \bm f_{t}$. The fact that these effects enter both \eqref{eq:e} and \eqref{eq:x} means that $\bm x_{i,t}$ is endogenous. Another implication is that $e_{i,t}$ and $\bm x_{i,t}$ need not be cross-sectionally independent as in Hansen (1999) but that they can be dependent.

It is convenient to introduce the following time-stacked versions of \eqref{eq:y}--\eqref{eq:x}:
\begin{align}
	\bm y_i &= \bm X_i\bm{\beta}+\bm H_i(\gamma)\bm \delta + \bm e_{i},\label{eq:y_i}\\
	\bm e_{i}&= \bm F \bm \mu_{i} +\bm \varepsilon_{i},\\
	\bm X_i&= \bm F \bm \Lambda_{i}+ \bm V_{i},\label{eq:x_i}
\end{align}
where $\bm y_i:=(y_{i,1},...,y_{i,T})'$, $\bm e_i:=(e_{i,1},...,e_{i,T})'$ and $\bm \varepsilon_{i} := (\varepsilon_{i,1},...,\varepsilon_{i,T})'$ are $T\times 1$ vectors, $\bm X_i:=(\bm x_{i,1},...,\bm x_{i,T})'$ and $\bm V_{i}:=(\bm v_{i,1}',...,\bm v_{i,T}')'$ are $T\times k$, $\bm H_i(\gamma):= (\bm h_{i,1}(\gamma) ,..., \bm h_{i,T}(\gamma))'$ is $T\times l$, and $\bm F:=(\bm f_{1},...,\bm f_{T})'$ is $T\times m$.

The parameters of interest are the slope coefficients $\bm{\beta}$ and $\bm \delta$, and the threshold parameter $\gamma$, the true values of which are henceforth denoted by $\bm{\beta}_0$, $\bm \delta_0$ and $\gamma_0$, respectively. The estimation of these parameters is challenging because the model is nonlinear in $\gamma_0$, and because of the endogeneity and cross-correlation induced by the interactive fixed effects. In order to describe the estimation approach, it is useful to first consider the case when $\gamma_0$ is known. With $\gamma_0$ known, \eqref{eq:y_i} is linear in $\bm{\beta}_0$ and $\bm \delta_0$, which means that the estimation can be carried out using the CCE approach of Pesaran (2006). Denote by $\bm{\bar{A}}=N^{-1}\sum_{i=1}^N \bm A_i$ the cross-section average of any generic variable $\bm A_i$. Standard implementation of CCE calls for the use of $\bm{\bar y}$ and $\bm{\bar X}$ as estimators of the space spanned by the factors in $\bm F$, and to ``defactor'' the data using these averages. However, in piecewise linear models, $\bm{\bar y}$ is not informative of $\bm F$ (see Karavias et al., 2023), and so we only use $\bm{\bar X}$. Let us therefore define $\bm Q_{\bm {\bar{X}}}:=\bm I_{T}-\bm {\bar{X}}(\bm{ \bar{X}}'\bm{ \bar{X}})^{-1}\bm {\bar{X}}'$ and denote by $\bm{\tilde{A}}_i := \bm Q_{\bm{\bar{X}}}\bm{A}_i$ the defactored version of any $T$-rowed variable $\bm{A}_i$. The defactored version of \eqref{eq:y_i} is given by
\begin{equation}\label{eq:tildemod}
	\bm{\tilde{y}}_i =\bm{ \tilde{X}}_i \bm{\beta}_0 + \bm {\tilde{H}}_{i}(\gamma_0)\bm \delta_0 + \bm {\tilde{e}}_{i} =\bm {\tilde W}_i(\gamma_0)\bm \theta_0+\bm{ \tilde{e}}_{i},
\end{equation}
where $\bm {\tilde W}_i(\gamma_0):=[\bm{ \tilde X}_i , \bm{\tilde H}_i(\gamma_0)]$ and $\bm{\theta}_0:=(\bm{\beta}_0',\bm{\delta}_0')'$ are $T\times (k+l)$ and $(k+l)\times 1$, respectively. This model can be stacked again, now over the cross-sectional units, giving
\begin{equation}\label{eq:ystack}
	\bm{\tilde{Y}}= \bm{\tilde{W}}(\gamma_0)\bm{\theta}_0 +\bm{\tilde{E}},
\end{equation}
where $\bm{\tilde{Y}}:=(\bm{\tilde{y}}_{1}',...,\bm{\tilde{y}}_{N}')'$ and $\bm{\tilde{E}}:=(\bm{\tilde{e}}_{1}',...,\bm{\tilde{e}}_{N}')'$ are $NT\times 1$, and $\bm{\tilde{W}}(\gamma_0):=(\bm{\tilde{W}}_{1}(\gamma_0)',...,\bm{\tilde{W}}_{N}(\gamma_0)')'$ is $NT\times (k+l)$. In this notation, the known-$\gamma_0$ CCE estimator of $\theta_0$ is given simply by:
\begin{equation}\label{eq:thetahat1}
	\bm{\tilde\theta}(\gamma_0):= (\bm{\tilde{W}}(\gamma_0)'\bm{\tilde{W}}(\gamma_0))^{-1}\bm{\tilde{W}}(\gamma_0)'\bm{\tilde Y}.
\end{equation}

If $\gamma_0$ is unknown, analogous to Hansen (1999, 2000), we propose minimizing the sum of squared CCE residuals over all possible values of $\gamma$;
\begin{equation}\label{eq:hatgamma}
	\hat \gamma := \arg\min_{ \gamma \in \Gamma_{NT}} SSR(\gamma),
\end{equation}
where
\begin{equation}\label{eq:ssrgamma}
	SSR(\gamma):=(\bm{\tilde Y}-\bm{\tilde W}(\gamma)\bm{\tilde\theta}(\gamma))'(\bm{\tilde Y}-\bm{\tilde W}(\gamma)\bm{\tilde\theta}(\gamma))
\end{equation}
with $\Gamma_{NT}:=\Gamma \cap \{z_{i,t}: i \in \{1,...,N\},\, t \in \{1,...,T\} \}$ and $\Gamma$ is a bounded set that contains $\gamma$. The minimization requires estimating \eqref{eq:y} and evaluating $SSR(\gamma)$ at most $NT$ times; once for each point in $\Gamma_{NT}$, which can be time-consuming. Hansen (1999, 2000) therefore proposes to approximate $\Gamma_{NT}$ with a grid of size $n < NT$. Denote by $z_{(j)}$ the $(j/n)$-th quantile of $\{z_{i,t}: i \in \{1,...,N\},\, t \in \{1,...,T\} \}$. The idea is to evaluate $SSR(\gamma)$ not at every point but at $\Gamma \cap\{z_{(1)},...,z_{(n)}\}$. The approximation can be made arbitrarily accurate by simply making the quantiles finer.

Given $\bm{\hat \gamma}$, we estimate $\bm{\theta}_0$ as
\begin{equation}\label{eq:theta_2}
	\bm{\hat\theta}: = \left[\begin{array}{c}
		\bm{\hat \beta} \\
		\bm{\hat \delta}
	\end{array}
	\right] : = \bm{\hat\theta}(\hat\gamma):=(\bm{\hat{W}}(\hat\gamma)'\bm{\hat{W}}(\hat\gamma))^{-1}\bm{\hat{W}}(\hat\gamma)'\bm{\hat Y},
\end{equation}
where $\bm{\hat{W}}(\hat\gamma)$ and $\bm{\hat Y}$ are $\bm{\tilde{W}}(\hat\gamma)$ and $\bm{\tilde Y}$, respectively, with the matrix $\bm{Q}_{\bm{\bar X}}$ replaced by $\bm{Q}_{\bm{\bar W}}:=\bm{I}_T-\bm{\bar W}\left(\bm{\bar W}'\bm{\bar W}\right)^{-1}\bm{\bar W}'$ with $\bm{\bar W}:=[\bm{\bar X}, \bm{\bar H}(\hat\gamma)]$. This last modification is necessary in order to asymptotically eliminate the interactive fixed effects.

So far we have presented the threshold estimator in \eqref{eq:hatgamma} and the slope coefficient estimator in \eqref{eq:theta_2}. We now focus on confidence intervals and hypothesis testing. As we demonstrate in Section \ref{sect:asy}, $\bm{\hat\theta}(\hat\gamma)$ supports asymptotically standard normal and chi-square inference as $N\to\infty$ with $T$ held fixed. However, this requires a consistent estimator of the asymptotic covariance matrix. The following estimator of the plug-in type does the trick:
\begin{equation}\label{eq:asyvar}
	\widehat{\mathrm{var}}(\bm{\hat\theta}):=\frac{1}{N}\bm{\hat\Omega}^{-1}\bm{\hat \Psi}\bm{\hat\Omega}^{-1}.
\end{equation}
Here $\bm{\hat\Omega} : = \bm{\hat\Omega}(\hat\gamma)$ and $\bm{\hat\Psi} :=\bm{\hat\Psi}(\hat\gamma)$, where
\begin{align}
	\bm{\hat\Omega}(\gamma) & :=\frac{1}{N}\bm{\hat W}(\gamma)'\bm{\hat W}(\gamma),\label{eq:Omega}\\
	\bm{\hat\Psi}(\gamma) & : =\frac{1}{N}\sum_{i=1}^{N} \bm{\hat W}_{i}(\gamma)'\bm{\hat\varepsilon}_{i}(\gamma)\bm{\hat\varepsilon}_{i}(\gamma)'\bm{\hat W}_{i}(\gamma), \label{eq:Psi}
\end{align}
with $\bm{\hat\varepsilon}_{i}(\gamma):=\bm{\hat y}_i-\bm{\hat W}_{i}(\gamma)\bm{\hat \theta}(\gamma)$ for $\gamma\in \Gamma$.

Denote by $\bm{R}$ a $r\times (k+l)$ matrix of rank $r\leq k+l$ and by $\bm q$ a $r\times 1$ vector. In order to test the null hypothesis $H_0: \bm{R}\bm{\theta}_0=\bm{q}$ when $r=1$, the following $t$-statistic can be used:
\begin{equation}\label{eq:t-test}
	t_{\theta_0} : =\frac{\bm{R}(\bm{\hat\theta}-\bm{\theta}_0)}{\sqrt{\bm{R}\widehat{\mathrm{var}}(\bm{\hat\theta})\bm{R}'}},
\end{equation}
whereas if $r\geq 1$, the following Wald statistic can be used:
\begin{equation}\label{eq:w-test}
	W_{\theta_0}: = [\bm{R}(\bm{\hat\theta}-\bm{\theta}_0)]'[\bm{R} \widehat{\mathrm{var}}(\bm{\hat\theta})\bm{R}']^{-1}[\bm{R}(\bm{\hat\theta}-\bm{\theta}_0)].
\end{equation}

The above test statistics can be used to test general hypotheses on the slope coefficients. Hypothesis testing on the threshold parameter of the form $H_0:\gamma=\gamma_0$ can be tested using the following likelihood ratio (LR) statistic:
\begin{equation}\label{eq:w}
	LR_{\gamma_0} : =N(T-k-l)\frac{SSR(\gamma_0)-SSR(\hat\gamma)}{SSR(\hat\gamma)}.
\end{equation}

Testing the above hypotheses is straightforward using the critical values provided in Section \ref{sect:asy}. Testing the null hypothesis of no threshold effects, $H_0:\bm \delta=\bm 0_{l\times 1}$, is more complicated in that $\gamma_0$ appears only under the alternative. The main exception is if $\gamma_0$ is known, in which case
\begin{equation}
	W_{\delta_0}(\gamma_0) = [\bm{R}\bm{\hat\theta}(\gamma_0)]'[\bm{R} \widehat{\mathrm{var}}(\bm{\hat\theta}(\gamma_0))\bm{R}']^{-1}[\bm{R}\bm{\hat\theta}(\gamma_0)]
\end{equation}
with $\bm{R} = (\bm{0}_{l\times k},\bm{I}_l)$ can be used and it would be asymptotically chi-squared distributed. If $\gamma_0$ is unknown, as it usually is, the following statistic of the sup type can be used:
\begin{equation}\label{eq:nothres}
	\mathrm{sup}W_{\delta_0} : =\sup_{\gamma\in \Gamma}W_{\delta_0}(\gamma)= W_{\delta_0}(\hat\gamma).
\end{equation}
However, now the asymptotic distribution is no longer standard, which requires bootstrapping, as pointed out by, for example, Hansen (1996), and Miao et al. (2020). The particular bootstrap that we employ here is a version of the pairs bootstrap of Kapetanios (2008), and De Vos and Stauskas (2024), which is not only simple to implement but also very general when it comes to the type of time series dependence that can be permitted. The steps are as follows:
\begin{enumerate}
	\item Generate the bootstrap sample $\{\bm{y}_i^*,\bm{X}_i^*\}_{i=1}^N$ by drawing pairs $(\bm{y}_i^*,\bm{X}_i^*)$ with replacement from the original sample $\{\bm{y}_i,\bm{X}_i\}_{i=1}^N$.
	
	\item Use $\{\bm{y}_i^*,\bm{X}_i^*\}_{i=1}^N$ to compute $\hat\gamma$, $\bm{\hat\theta}$ and $\widehat{\mathrm{var}}(\bm{\hat\theta})$. Denote these estimates as $\hat\gamma^*$, $\bm{\hat\theta}^* : = \bm{\hat\theta}^*(\hat\gamma^*)$ and $\widehat{\mathrm{var}}^*(\bm{\hat\theta}^*)$, respectively. Then calculate
	\begin{equation}
		\mathrm{sup}W_{\delta_0,1}^* := [\bm{R}(\bm{\hat\theta}^*-\bm{\hat\theta})]'[\bm{R} \widehat{\mathrm{var}}^*(\bm{\hat\theta}^*)\bm{R}']^{-1}[\bm{R}(\bm{\hat\theta}^*-\bm{\hat\theta})].
	\end{equation}
	
	\item Repeat steps 1 and 2 $B-1$ times to obtain $\mathrm{sup}W_{\delta_0,2}^*,...,\mathrm{sup}W_{\delta_0,B}^*$.
	
	\item Calculate the bootstrap $p$-value in the following way:
	\begin{equation}
		p := \frac{1}{B}\sum_{b=1}^{B}\mathbb{I}(\mathrm{sup}W_{\delta_0,b}^*>\mathrm{sup}W_{\delta_0}).
	\end{equation}
	Reject $H_0: \bm{\delta}_0 = \bm 0_{l\times 1}$ if $p$ is less than the required significance level $\phi$.
\end{enumerate}

This completes the new toolbox. It contains; (i) the threshold estimator $\hat \gamma$ in \eqref{eq:hatgamma}, (ii) the $LR_{\gamma_0}$ test in \eqref{eq:w} for testing hypotheses regarding $\gamma_0$, (iii) the slope estimator $\bm{\hat\theta}$ in \eqref{eq:theta_2} and its estimated covariance matrix in \eqref{eq:asyvar}, which can be used for confidence interval and test construction as in \eqref{eq:t-test} and \eqref{eq:w-test}, and (iv) the $\mathrm{sup}W_{\delta_0}$ test in \eqref{eq:nothres} for testing the null hypothesis of no threshold effect.

\section{Asymptotic results}\label{sect:asy}

\subsection{Assumptions}

We begin this section by introducing some notation. If $\bm{A}$ is a matrix, $\mathrm{rank}(\bm{A})$, $\mathrm{tr}(\bm{A})$, $\|\bm{A}\| : = \sqrt{\mathrm{tr}(\bm{A}'\bm{A})}$, $\lambda_{\min}(\bm{A})$ and $\lambda_{\max}(\bm{A})$ signify its rank, trace, Frobenius norm, and smallest and largest eigenvalues, respectively. The symbols $\to_{d}$ and $\to_p$ signify convergence in distribution and convergence in probability, respectively. We use w.p.1 (w.p.a.1) to denote with probability (approaching) one.

\begin{assumption}[Time dimension]\label{ass:t}\
	$T>k+l$.
\end{assumption}

\begin{assumption}[Factors]\label{ass:f}\
	$\bm F$ is non-random such that	$\mathrm{rank}(\bm{F})=m$.
\end{assumption}

\begin{assumption}[Loadings]\label{ass:load}\
	$\bm{\Lambda}_i$ is non-random such that $\sup_{i}\|\bm{\Lambda}_i\|<\infty$, $\mathrm{rank}(\bm{\bar \Lambda}) = m\leq k$ for all $N < \infty$, and $\bm{\bar \Lambda} \to \bm{\Lambda}$ as $N\to\infty$, where $\mathrm{rank}(\bm{\Lambda}) = m\leq k$ and $\|\bm{\Lambda}\|<\infty$.
\end{assumption}

The results reported in this section do not require that $T$ is large, provided that Assumptions \ref{ass:t} is met. This is an advantage in the sense that in practice $T$ is always finite. However, we do require $N\to\infty$. This means that the asymptotic approximation offered by our results is expected to work well in ``short'' panels where $N$ is larger than $T$. The main drawback is that our theory is silent about the effect of increasing $T$. However, intuition suggests that the proposed toolbox should be valid also when $T\to\infty$, provided that suitable regulatory conditions are met, and our Monte Carlo results confirm this.

CCE studies standardly assume that both factors and loadings follow certain probability laws (see Pesaran, 2006). Assumptions \ref{ass:f} and \ref{ass:load} treat these objects as non-random, which is a more general consideration, since it is non-parametric. Assumption \ref{ass:load} requires that the number of factors must not be larger than the number of regressors used to estimate them.\footnote{See Juodis et al. (2021) for a discussion and some results for the case when Assumption \ref{ass:load} fails.} One way to relax this condition is if some of the factors are observed, because Assumption \ref{ass:load} only applies to the loadings of unobserved factors. Observed factors like a unit-specific fixed effects (a vector of ones) can be appended to $\bm{F}$ in Assumption \ref{ass:f} and to $\bm{\bar X}$ in the estimation. The rest is unaffected. In Section \ref{sect:appl}, we elaborate on this point and explain how to test Assumption \ref{ass:load}.

\begin{assumption}[Errors]\label{ass:err}
	$\varepsilon_{i,t}$ and $\bm{v}_{i,t}$ are independent of each other as well as across $i$ with $\mathbb{E}(\varepsilon_{i,t})=0$, $\sigma_{\varepsilon}^2: =\lim_{N\to\infty}N^{-1}\sum_{i=1}^N\sum_{t=1}^T\mathbb{E}( \varepsilon_{i,t}^2 ) > 0$, $\inf_{i,t}\lambda_{\min}(\mathbb{E}(\bm{v}_{i,t}\bm{v}_{i,t}'))>0$, $\sup_{i,t}\mathbb{E}(\varepsilon_{i,t}^4) < \infty$ and $\sup_{i,t}\mathbb{E}(\|\bm{v}_{i,t}\|^4)<\infty$.
\end{assumption}

Assumption \ref{ass:err} is very general in that it puts no restrictions on the allowable serial correlation and heteroskedasticity properties of the errors. The condition that $\varepsilon_{i,t}$ and $\bm{v}_{i,t}$ are independent of each other rules out the presence of lagged dependent variables in $\bm{x}_{i,t}$. This can be relaxed but then $\varepsilon_{i,t}$ must be serially uncorrelated.

Our next assumption is stated in terms of the following matrices:
\begin{align}
	\bm{\Omega}(\gamma) & :=\lim_{N\to\infty}\frac{1}{N}\sum_{i=1}^N \mathbb{E}[ \bm{W}_i(\gamma)'\bm{Q}_{\bm{F}}\bm{W}_i(\gamma)],\\
	\bm{\Gamma}(\gamma_1,\gamma_2) & :=\underset{N\to\infty}{\mathrm{plim}}\, \bm{\Gamma}_N(\gamma_1,\gamma_2),
\end{align}
where
\begin{align}
	\bm{\Gamma}_N(\gamma_1,\gamma_2) & := \frac{1}{N}\sum_{i=1}^N \bm{H}_i(\gamma_1)'\bm{Q}_{\bm{F}}\bm{H}_i(\gamma_2) \notag\\
	& - \frac{1}{N}\sum_{i=1}^N\bm{H}_i(\gamma_1)'\bm{Q}_{\bm{F}}\bm{X}_i\left(\frac{1}{N}\sum_{i=1}^N \bm{X}'\bm{Q}_{\bm{F}}\bm{X}_i\right)^{-1}\frac{1}{N}\sum_{i=1}^N \bm{X}'\bm{Q}_{\bm{F}}\bm{H}_i(\gamma_2)
\end{align}
with $\gamma, \gamma_1, \gamma_2\in \Gamma$.

\begin{assumption}[Identification]\label{ass:invrt}\
	\begin{enumerate}
		\item[(a)] $\inf_{\gamma \in \Gamma}\lambda_{\min}(\bm{\Omega}(\gamma)) > 0$ and $\inf_{\gamma \in \Gamma}\lambda_{\min}(\bm{\Gamma}(\gamma,\gamma)) > 0$ for all $T$.
		
		\item[(b)] $\inf_{\gamma}\lambda_{\min}(\bm{\Sigma}(\gamma)) > 0$ for all $T$, where
		\begin{equation*}
			\bm{\Sigma}(\gamma):=\bm{\Gamma}(\gamma_0,\gamma_0)-\bm{\Gamma}(\gamma_0,\gamma)[\bm{\Gamma}(\gamma,\gamma)]^{-1}\bm{\Gamma}(\gamma,\gamma_0).
		\end{equation*}
	\end{enumerate}
\end{assumption}

Assumption \ref{ass:invrt} is a non-collinearity condition that ensures identification of both $\bm \theta_0$ and $\gamma_0$.

\begin{assumption}[Shrinking threshold]\label{ass:shrink}\
	$\bm{\delta}_0= N^{-\alpha} \bm{c}_0$ with $\bm{c}_0\neq 0_{l\times 1}$ and $\alpha\in (0,1/2)$. 		
\end{assumption}

The shrinking threshold condition in Assumption \ref{ass:shrink} is standard in the threshold effects literature (see, for example, Hansen, 1999, 2000, and Miao et al., 2020). It is needed to show that the asymptotic distributions of $\hat\gamma$, $\bm{\hat \beta}$ and $\bm{\hat\delta}$ are nuisance parameter free, and cannot be dispensed with without affecting the results.  Our asymptotic distributions are therefore expected to provide a better approximation to the true sampling distributions when the threshold effect is relatively small. If the threshold effect is large, while the threshold parameter will be accurately estimated, inference based on Assumption \ref{ass:shrink} may be misleading.

Before we come to our next assumption, we again introduce some notation. In particular, let us define
\begin{align}
	\bm{\Xi}_N(\gamma) &:=\frac{1}{\sqrt{N}}\sum_{i=1}^N  \bm{H}_i(\gamma)'\bm{Q}_{\bm{F}} \bm{\varepsilon}_i,\\
	\bm{\Phi}_{N}(\gamma) & : = \frac{1}{N}\sum_{i=1}^N\bm{\Delta H_i}(\gamma)'\bm{Q}_{\bm{F}}\bm{\Delta H}_i(\gamma),\\
	\bm{\Theta}_{N}(\gamma) & : = \frac{1}{N}\sum_{i=1}^N\bm{\Delta  H}_i(\gamma)'\bm{Q}_{\bm{F}} \bm{\varepsilon}_i \bm{\varepsilon}_i'\bm{Q}_{\bm{F}}\bm{\Delta  H}_i(\gamma),
\end{align}
where $\bm{\Delta H}_i(\gamma):= \bm{H}_i(\gamma)-\bm{H}_i(\gamma_0)$. The limits of the last two matrices are given by $\bm{\Phi}(\gamma):=\lim_{N\to\infty}\mathbb{E}(\bm{\Phi}_{N}(\gamma))$ and $\bm{\Theta}(\gamma):=\lim_{N\to\infty}\mathbb{E}(\bm{\Theta}_{N}(\gamma))$, respectively.

\begin{assumption}[High-order moments]\label{ass:hom}\			
	\begin{enumerate}
		\item[(a)] $\bm{\Phi}(\gamma)$ and $\bm{\Theta}(\gamma)$ are continuous at $\gamma=\gamma_0$;
		
		\item[(b)] Define $a_N: = N^{1-2\alpha}$. There are constants $B>0$ and  $d_1, d_2 \in (0,\infty)$ such that for all $\eta>0$ and $\epsilon>0$, there exists a $\bar \mu< \infty$, such that for large enough $N$ and all $T$
		\begin{align*}
			& \mathbb{P}\Bigg(\inf_{a_N^{-1}\bar \mu\leq|\gamma-\gamma_0|<B}\frac{ \bm c_0'\bm{\Phi}_{N}(\gamma)\bm c_0}{|\gamma-\gamma_0|} < (1-\eta)d_1 \Bigg)\leq\epsilon; \\
			& \mathbb{P}\Bigg(\sup_{a_N^{-1}\bar \mu \leq|\gamma-\gamma_0|<B}\frac{ \|\bm{\Phi}_{N}(\gamma)\| }{|\gamma-\gamma_0|} > (1+\eta)d_2 \Bigg)\leq\epsilon; \\
			& \mathbb{P}\Bigg(\sup_{a_N^{-1}\bar \mu\leq|\gamma-\gamma_0|<B}\frac{ \|\bm{\Xi}_{N}(\gamma)-\bm{\Xi}_{N}(\gamma_0)\| }{|\gamma-\gamma_0|} >\eta \Bigg)\leq\epsilon.		
		\end{align*}
		
		\item[(c)] $\inf_\gamma \bm{c}_0'\bm{\Phi}(\gamma)\bm{c}_0 > 0$ and $\sup_\gamma \bm{c}_0'\bm{\Theta}(\gamma)\bm{c}_0 < \infty$ for all $T$.
		
		\item[(d)] $N^{-\alpha}\sum_{i=1}^N \bm{c}_0'\bm{\Delta  H}_i(\gamma)'\bm{Q}_{\bm{F}} \bm{\varepsilon}_i \to_d \sqrt{\bm{c}_0'\bm{\Theta}(\gamma_0)\bm{c}_0}B(\gamma)$ as $N\to\infty$, where $B(\gamma)$ is a standard two-sided Brownian motion.
	\end{enumerate}
\end{assumption}

Assumption \ref{ass:hom} is a high-level moment condition that is needed to establish the rate of consistency and asymptotic distribution of $\hat \gamma$. The assumption is stated directly in terms of the required probability bounds and convergence results. See Hansen (1999, 2000), and Miao et al. (2020) for more primitive sufficient conditions that give rise to Assumption \ref{ass:hom}.

\subsection{Results}

\begin{theorem}\label{th:consrate}
	Suppose that Assumptions \ref{ass:t}--\ref{ass:hom} are met. Then, the following results hold as $N\to \infty$:
	\begin{itemize}
		\item[(a)] $N^{1-2\alpha}(\hat\gamma-\gamma_0)=O_p(1)$;
		\item[(b)] $\sqrt{N}(\bm{\hat\theta}-\bm{\theta}_0) = O_p(1)$.
	\end{itemize}
\end{theorem}

Consider part (a) of the theorem. Recall from Assumption \ref{ass:shrink} that $\alpha\in (0,1/2)$. Hence, $\hat\gamma-\gamma_0 = O_p(N^{2\alpha-1}) = o_p(1)$ and therefore $\hat\gamma$ is consistent for $\gamma_0$, and for $\alpha\in (0,1/4)$ it is even superconsistent, which is in line with the results of Hansen (2000), and Miao et al. (2020). We also see that the rate of convergence is faster the closer is $\alpha$ to zero, which is just as expected because $\alpha$ measures the rate of shrinking of the threshold effect, $\bm{\delta}_0$. Hence, by making $\alpha$ smaller, $\bm{\delta}_0$ becomes larger making it easier to discern. The fact that $\hat\gamma$ may be superconsistent is a key strength of the discontinuous threshold model considered here, as opposed to alternatives such as the smoothed threshold model of Seo and Linton (2007), and the kinked threshold model of Hansen (2017). But while $\hat\gamma$ can be superconsistent, $\hat\theta$ cannot, as is clear from Theorem 3.1 (b).

A word on the role of Assumption \ref{ass:shrink} in Theorem \ref{th:consrate}: Assumption \ref{ass:shrink} is only needed in order to derive the rate of convergence of $\hat\gamma$ in part (a), and is not required for part (b). Hence, $\bm{\hat\theta}$ is $\sqrt{N}$-consistent even if $\bm{\delta}_0$ does not shrink to zero. In fact, in contrast to what Miao et al. (2020) claim, $\hat\gamma$ is also consistent when $\bm{\delta}_0$ does not shrink to zero (see the online appendix for a proof), even if the rate of convergence cannot be established without Assumption \ref{ass:shrink}.

\begin{theorem}\label{th:asydist1}
	Suppose that Assumptions \ref{ass:t}--\ref{ass:hom} are met. Then, the following results hold as $N\to \infty$:
	\begin{itemize}
		\item[(a)] $\sqrt{N}(\bm{\hat\theta}-\bm{\theta}_0)\to_{d} N(\bm{0}_{(k+l)\times 1},\mathrm{var}(\bm{\hat\theta}))$, where $\mathrm{var}(\bm{\hat\theta}):= \bm{\Omega}^{-1}\bm{\Psi}\bm{\Omega}^{-1}$ with $\bm{\Omega} := \lim_{N\to\infty}E(\bm{\hat\Omega})$ and $\bm{\Psi} := \lim_{N\to\infty}E(\bm{\hat\Psi})$;
		
		\item[(b)] $N^{1-2\alpha}(\hat\gamma-\gamma_0)\to_{d}\omega \zeta$, where the parameter  $\omega:=\bm{c}_0'\bm{\Theta}(\gamma_0)\bm{c}_0/(\bm{c}_0'\bm{\Phi}(\gamma_0)\bm{c}_0)^2$ and the distribution $\zeta:=\arg\max_{r\in (-\infty,\infty)}[-|r|/2+B(r)]$, with $B(r)$ being a two-sided standard Brownian motion.
	\end{itemize}	
\end{theorem}

The asymptotic distributions reported in Theorem \ref{th:asydist1} are similar to those reported by Miao et al. (2020). Because $N\widehat{\mathrm{var}}(\bm{\hat\theta}) \to_p \mathrm{var}(\bm{\hat\theta})$ as $N\to\infty$, part (a) implies that $t_{\theta_0}$ and $W_{\theta_0}$ will be asymptotically $N(0,1)$ and $\chi^2(r)$, respectively, under the null hypothesis. Wald and $t$-test inference for $\gamma_0$ based on part (b) is made complicated by the fact that $\omega$ depends on the unknown $\gamma_0$, which is known to lead to poor small-sample performance (see Hansen, 2000, for a discussion). In this paper, we therefore follow Hansen (1999, 2000) and Miao et al. (2020) and focus instead on the LR test, which is more robust in this regard.

\begin{theorem}\label{th:asydistlr}
	Suppose that Assumptions \ref{ass:t}--\ref{ass:hom} are met and that $H_0:\gamma=\gamma_0$ holds. Then, as $N\to \infty$,
	\begin{equation*}
		LR_{\gamma_0}\to_{d}\eta^2 \xi,
	\end{equation*}
	where $\eta^2:=\bm{c}_0'\bm{\Theta}(\gamma_0)\bm{c}_0/[\sigma_{\varepsilon}^2\bm{c}_0'\bm{\Phi}(\gamma_0)\bm{c}_0]$ and $\xi:=\sup_{r\in (-\infty,\infty)}[2B(r)-|r|]$.
\end{theorem}

As Hansen (1999) shows, the cumulative distribution function (CDF) of $\xi$ is given by $P(\xi \leq r)=(1-\exp(-r/2))^2$, which can be inverted to obtain the following quantile function: $c(\phi):=-2\log(1-\sqrt{1-\phi})$, where $\phi$ is the significance level. Critical values at the 10\%, 5\% and 1\% levels are given by $c(0.1) = 5.94$, $c(0.05) = 7.35$ and $c(0.01) = 10.59$, respectively. These are suitable if $\varepsilon_{i,t}$ is homoskedastic, in which case $\bm{\Theta}(\gamma_0) = \sigma_{\varepsilon}^2\bm{\Phi}(\gamma_0)$ so that $\eta^2 = 1$. The decision rule in this case is to reject $H_0$ if $LR_{\gamma_0} > c(\phi)$. Confidence intervals for $\gamma_0$ can be constructed by inverting the LR test statistic itself (see Hansen, 2000). The appropriate $100(1-\phi)$\% confidence interval is given by
\begin{equation}
	CI_{\gamma_0}(\phi): = \{\gamma\in \Gamma: 	LR_{\gamma_0} \leq c(\phi)\}.
\end{equation}
If $\varepsilon_{i,t}$ is heteroskedastic, the appropriate test statistic to consider is not $LR_{\gamma_0}$ but $LR_{\gamma_0} / \hat \eta^2$, where $\hat \eta^2$ is a consistent estimator of $\eta^2$. Analogous to Miao et al. (2020), we use
\begin{equation}
	\hat\eta^2:=N\frac{\hat\delta'[\sum_{i=1}^N\sum_{t=1}^T K_b(\hat\gamma-z_{i,t})\bm{x}_{i,t}\bm{x}_{i,t}'\hat\varepsilon_{i,t}^2]\hat\delta}{SSR(\hat\gamma)\hat\delta'[\sum_{i=1}^N\sum_{t=1}^T K_b(\hat\gamma-z_{i,t}) \bm{x}_{i,t}\bm{x}_{i,t}']\hat\delta},
\end{equation}
where $K_b(s)=b^{-1}K(s/b)$, $K(\cdot)$ is a kernel function and $b$ is the bandwidth parameter satisfying $b\to 0$.

\begin{theorem}\label{th:asydist}
	Suppose that Assumptions \ref{ass:t}--\ref{ass:invrt} and \ref{ass:hom} are met, and that $H_0:\bm{\delta}_0=\bm{0}_{l\times 1}$ holds. Then, as $N\to \infty$,
	\begin{equation*}
		\mathrm{sup}W_{\delta_0} \to_{d} \sup_{\gamma\in\Gamma} \vartheta(\gamma),
	\end{equation*}
	where $\vartheta(\gamma):=\bm{G}(\gamma)'\bm{\Omega}(\gamma)^{-1}\bm{R}'[\bm{R}\bm{\Omega}(\gamma)^{-1}\bm{\Psi}(\gamma)\bm{\Omega}(\gamma)^{-1}\bm{R}']^{-1}\bm{R}\bm{\Omega}(\gamma)^{-1}\bm{G}(\gamma)$ with
	$\bm{G}(\gamma)$ being a certain mean zero normal $(k+l)\times 1$ vector defined in the appendix such that $\bm{R}\bm{\Omega}(\gamma)^{-1}\bm{G}(\gamma)$ has covariance kernel $\bm{R}\bm{\Omega}(\gamma)^{-1}\bm{\Psi}(\gamma)\bm{\Omega}(\gamma)^{-1}\bm{R}'$.
\end{theorem}

As Theorem \ref{th:asydist} makes clear, $\vartheta(\gamma)$ is not nuisance parameter free, which invalidates standard inference. However, the bootstrapped version of $\mathrm{sup}W_{\delta_0}$ converges to the same asymptotic distribution, suggesting that the bootstrap should enable asymptotically valid inference.\footnote{The basic intuition for why the bootstrap should work is that with $\gamma_0$ known, the model is linear as pointed out in Section \ref{sect:mod}, and then validity should follow from arguments similar to those of De Vos and Stauskas (2024). Of course, in our paper $\gamma_0$ is unknown; however, we can estimate it consistently. This means that asymptotically observing $\hat \gamma$ is just as good as observing $\gamma_0$ itself. This intuition is supported by our Monte Carlo results.}

\section{Monte Carlo results}\label{sect:mc}

A small-scaled Monte Carlo simulation exercise was undertaken to investigate the small-sample properties of the proposed toolbox. The data generating process used for this purpose is given by a restricted version of equations (2.1)--(2.3) in the main text that sets $k=2$, $m=l=1$ and $\bm{\beta}_0 = (\beta_{0,1},\beta_{0,2})' = (1,2)'$. Also, $\bm{S} = (0,1)'$ and $z_{i,t} = x_{2,i,t}$, such that $h_{i,t}(\gamma) = \bm{S}'\bm x_{i,t}\mathbb{I}(x_{2,i,t}>\gamma)= x_{2,i,t}\mathbb{I}(x_{2,i,t}>\gamma)$. Several specifications of $\bm{\delta}_0$ are considered. The loading $\mu_{i}$ is drawn from $N([1,2]',0.2I_2)$, while $\bm{\Lambda}_i = \bm{I}_2 + \bm{u}_i$, where the elements of $\bm{u}_i$ are all drawn from $N(0,0.5)$. This means that Assumption 3.3 is met. The elements of $\bm{v}_{i,t}$ are drawn from $N(0,1)$. Both factors and regression errors are allowed to be serial correlated;
\begin{align}
	f_{t}&=0.3 f_{t-1} + \eta_{t}, \\
	\varepsilon_{i,t}&=0.5 \varepsilon_{i,t-1}+\epsilon_{i,t},
\end{align}
where $\epsilon_{i,t} \sim N(0,1-0.5^2)$ and $\eta_{t} \sim N(0,1-0.3^2)$. We generate $2,000$ samples of size $N\in \{25,50,100,200\}$ and $T\in \{10,25,50,100\}$. Hence, not all values of $T$ ($N$) are small (large).

\begin{center}
	{\sc Insert Tables 1--3 about here}
\end{center}

The results are reported in Tables 1--3. We begin by considering the results contained in Table 1 on the empirical rejection frequencies of a nominal 5\% level $\mathrm{sup}W_{\delta_0}$ test based on $B=300$ bootstrap repetitions. While when $\bm \delta_0=0$ these frequencies represent size, when $\bm \delta_0 \in \{0.25, 0.5\}$ they represent power. The size is quite close to the nominal level for all constellations of $N$ and $T$ considered. The test is slightly oversized when $N$ is small; however, this effect quickly disappears as $N$ increases. As expected, power in increasing not only in the size of the threshold effect, as measured by $\bm \delta_0$, but also in the size of the sample, as measured by $N$ and $T$.

Consider next the results reported in Table 2 on the small-sample properties of $\hat\gamma$. We report the bias, the root mean squared error (RMSE) and the empirical coverage of a 95\% confidence interval for $\gamma_0$. In contrast to before, now we set $\bm \delta_0 =N^{-\alpha}$ with $\alpha\in\{0,0.1,0.2\}$. Hence, unless $\alpha=0$ so that $\bm \delta_0 =1$ is fixed, the threshold is shrinking, as required by Assumption 3.6. The performance in terms of bias and RMSE is increasing in $N$ and $T$, which is partially expected given the consistency of $\hat\gamma$ as $N\to \infty$. Performance decreases as $\alpha$ grows, which is consistent with a smaller value of $\gamma_0$ being more difficult to estimate. We also see that the effect of increasing $N$ and $T$ is more pronounced the smaller is $\alpha$, which is consistent with the rate of convergence of $\hat\gamma$ being decreasing in $\alpha$ (Theorem 3.1). The empirical coverage is generally close to the nominal 95\% level. The main exception is when $\alpha$ is close to zero and $T$ is large, in which case coverage can be substantially lower than 95\%, which is partly expected given the discussion of Assumption 3.6.

Let us finally consider the 5\% size and power results reported in Table 3 for the $t_{\theta_0}$ test when testing $H_0: \beta_{0,1}=1$ against $H_1: \beta_{0,1}=0.9$, and $H_0: \bm \delta_0 =N^{-\alpha}$ against and $H_1: \bm \delta_0 =N^{-\alpha}-0.1$. The size of both tests is close to the nominal level for all combinations of $N$ and $T$, and for all values of $\alpha$. We also see that power increases with increasing values of both $N$ and $T$, and that it does so regardless of the value of $\alpha$. This suggests that Assumption 3.6 is not particularly important for inference regarding $\bm \theta_0$.

The overall conclusion is that the consistency and asymptotic distribution theories laid out in Section \ref{sect:asy} provide accurate approximations in small samples.

\section{Application}\label{sect:appl}

The literature concerned with the relationship between inflation and economic growth was pioneered by studies such as Fisher (1993), De Gregorio (1993) and Barro (1995). This first wave of papers found the relationship to be negative. However, this result has since then been overturned by evidence based on threshold regression. The accumulated empirical evidence can be summarized as follows: While some inflation is good for economic growth, too much is harmful. Many papers have tried to estimate the threshold (see Azam and Khan, 2022, and Nell, 2023, for recent overviews). Our point of entry is the highly cited paper of Khan and Senhadji (2001), which was among the first to apply Hansen's (1999) threshold regression approach in the present context.\footnote{Hansen's (1999) approach has since then become one of the workhorses of the industry (see Nell, 2023, for an overview).} Using panel data covering 140 countries between 1960 and 1998 they estimate the threshold to 1--3\% for industrialized economies and 11–-12\% for developing countries, estimates that have since then become benchmarks in the literature.

Of course, the results of Khan and Senhadji (2001) do not account for the recent surge in inflation following the COVID-19 pandemic, nor do they account for the fact that economic growth has been below trend ever since the global financial crisis. These recent developments may well have affected the inflation--economic growth relationship. There is also the role of government spending in this relationship, which has attracted considerable interest since the pandemic and the expansionary fiscal and monetary policies that followed. There is quite some evidence to suggest that government spending affects economic growth nonlinearly (see, for example, Asimakopoulos and Karavias, 2016). Khan and Senhadji (2001) ignore this effect, which is possible if inflation and government spending are unrelated. However, high government spending may well enable households and firms to spend more than they otherwise would, thus causing inflation to increase. If this is the case, ignoring government spending will lead to omitted variables bias and as a result misleading estimates of the inflation--economic growth relationship. Another source of concern is that Hansen's (1999) approach supposes that the cross-sectional units are independent, which is likely violated due to strong cross-country linkages.

The above observations suggest that there is a need to reevaluate the existing empirical evidence using the most recent data that include government spending and econometric techniques that allow for cross-sectional dependence. In the present section, we do exactly that. Our sample is collected from the World Bank Development indicators database. It covers 74 countries between 1970 and 2022, among which 51 (23) are developing (industrialized).\footnote{The included countries are Algeria, Australia, Austria, Bahamas, Bangladesh, Barbados, Belgium, Bolivia, Botswana, Burkina Faso, Burundi, Cameroon, Canada, Colombia, Costa Rica, Cote d'Ivoire, Cyprus, Denmark, Dominican Republic, Ecuador, Egypt, El Salvador, Fiji, Finland, France, Gabon Gambia, Germany, Ghana, Greece, Guatemala, Honduras, Hong Kong SAR, India, Indonesia, Iran, Ireland, Italy, Jamaica, Japan, Jordan, Kenya, Korea Rep, Luxembourg, Madagascar, Malaysia, Malta, Mauritius, Mexico, Morocco, Nepal, Netherlands, New Zealand, Niger, Norway, Pakistan, Panama, Paraguay, Portugal, Rwanda, Saudi Arabia, Senegal, Singapore, South Africa, Spain, Sri Lanka, Sudan, Sweden, Thailand, Togo, Tunisia, United Kingdom, United States, and Uruguay.} Following the convention in the literature (see, for example, Khan and Senhadji, 2001), the dependent variable ($y_{i,t}$) is GDP per capita growth (GDP), while the regressors ($x_{i,t}$) are inflation (INF), which is also the threshold variable ($z_{i,t}$), government spending (GOV), trade openness (TRA), population growth (POP) and gross capital formation (CAP).\footnote{Following the convention in the literature, INF is computed using the so-called ``partial log'' transformation that keeps values less than one as they are but transforms all other values into logs (see, for example, Ghosh and Phillips, 1998, Khan and Senhadji, 2001, Kremer et al., 2013, and Sarel, 1996).} Table 4 reports some descriptive statistics for these variables for the full sample as well as for the developing and industrialized subsamples. One of the descriptives is the CD test of Pesaran (2021), which tests the null hypothesis of no remaining cross-sectional correlation after controlling for country fixed effects. The null is firmly rejected at all conventional significance levels for all variables, suggesting that, as expected given the discussion of Section \ref{sect:intro}, country fixed effects do not address the issue of cross-country correlation.

\begin{center}
	{\sc Insert Tables 4 and 5 about here}
\end{center}

The results reported in Table 4 motivate our interactive fixed effects specification. As pointed out in Section \ref{sect:asy}, one way to relax Assumption \ref{ass:load} is if there are known factors. By treating a country-specific constant and normalized time trend as known factors, we can estimate more unknown factors. In order to also entertain the possibility that the impact of INF and GOV on GDP is not linear, we allow the coefficients of both regressors to be subject to threshold effects.

The results are reported in Table 5. Separate regressions are run for the full, developing and industrialized samples. The $\mathrm{sup}W_{\delta_0}$ test rejects the null hypothesis of no threshold effects ($H_0: \bm \delta_0 = 0$) for the full and developing country samples but not for the sample of industrialized countries, which could be due to the relatively small value of $N$ in this case.\footnote{The grid search is done by splitting the range of INF into 400 points and then trimming 5\% from the beginning and end.} The threshold parameter ($\gamma_0$) is estimated at $3.401$ for the industrialized countries, which means that the coefficients of INF and GOV break if inflation goes above $3.401\%$. The estimated 95\% confidence interval is given by $[2.787,4.336]$, which is quite narrow. This interval is close not only to the 1--3\% estimates reported for Khan and Senhadji (2001), and Kremer et al. (2013), but also to the inflation targets of most low-inflation industrialized countries (see Nell, 2023). In the high inflation regime the coefficient on INF is estimated to be significantly negative, while in the low inflation regime said coefficient is estimated to be positive, although not significantly so. We also see that the estimated effect is relatively larger in absolute value in the high inflation regime, suggesting that the ``price'' of high inflation is larger than the ``reward'' of low inflation (see Ghosh and Phillips, 1998, and Sarel, 1996, for similar findings). We interpret these results as providing support in favor of the conventional wisdom that inflation targets in industrialized countries should be kept in the 0--3\% range (see, for example, Burdekin et al., 2004). The effect of GOV is quite different from that of INF. In particular, while significantly negative in the low inflation regime, the coefficient on GOV is estimated to be significantly positive in the high inflation regime. This suggests that in bad times when inflation is high governments can stimulate growth by increasing public spending, which makes sense. The above picture is largely the same for the full sample of countries, except that the coefficient on INF is significant in the low inflation regime.

For developing countries the estimated threshold is $1.049\%$ with a 95\% confidence interval of $[1.029,1.306]$, which is at the low end of the scale when compared to the existing literature (see Azam and Khan, 2022, and Nell, 2023).\footnote{The 2.5\% threshold reported by Ghosh and Phillips (1998) is close but they do not differentiate between developing and industrialized countries, which means that their results are not directly comparable. It also raises the issue of aggregation bias.} We also note that this estimate is lower than that for industrialized countries, which is again unlike most existing studies. We do not claim to have the full explanation. However, we note that there is evidence to suggest that controlling for common observable country characteristics such as institutional quality leads to lower threshold estimates for developing countries (see, for example, Ibarra and Trupkin, 2016). Our interactive fixed effects specification is designed to address concerns like this. It provides a means to control for a wide range of common characteristics without for that matter requiring that they are observed. In order to test if the included cross-sectional averages together with the observed factors are enough to capture the factors present in the data, we apply the rank condition classifier of De Vos et al. (2024), which assesses Assumption \ref{ass:load}. The results reported in Table 5 provide no evidence against the assumption. We also apply Juodis and Reese's (2022) version of the CD test, which tests for remaining cross-sectional correlation in the regression residuals.\footnote{The methods of De Vos et al. (2024) and Juodis and Reese's (2022) are designed for linear panel data models but we conjecture that they are applicable here as well. The reason is that the model is linear for a given $\gamma_0$, as pointed out in Sections \ref{sect:mod} and \ref{sect:asy}.} While for industrialized countries the null hypothesis is marginally rejected at the 10\% level, at the 5\% level or better there is no evidence against the null for any of the samples, which reinforces the results based on the rank condition classifier.

\section{Conclusions}\label{sect:concl}

This paper develops a new toolbox for threshold regression that is the first to accommodate fixed-$T$ panel data in the presence of interactive fixed effects. It is also user-friendly and computationally efficient. The new toolbox is applied to reevaluate the inflation--growth relationship using the most recent data available. The main result is that this relationship is nonlinear with a threshold at 1.1\%--3.5\%, depending on the sample considered.

\pagebreak

\newpage

\begin{table}[h]
	\caption{5\% size and power results for $\mathrm{sup}W_{\bm \delta_0}$.}
	\hskip 12 pt
	\centering
	\footnotesize
	\par
	\begin{tabular}{@{}llllllllllllllll@{}}
		\toprule
		&  & \multicolumn{4}{l}{$\bm{\delta}_0=0$ (size)} &  & \multicolumn{4}{l}{$\bm{\delta}_0=0.25$ (power)} &  & \multicolumn{4}{l}{$\bm{\delta}_0=0.5$ (power)} \\ \cmidrule(lr){3-6} \cmidrule(lr){8-11} \cmidrule(l){13-16}
		$N$/$T$ &  & 10       & 25       & 50      & 100     &  & 10        & 25        & 50       & 100      &  & 10        & 25       & 50       & 100      \\ \hline
		&  &          &          &         &         &  &           &           &          &          &  &           &          &          &          \\
		25  &  & 0.060    & 0.065    & 0.055   & 0.041   &  & 0.135     & 0.378     & 0.719    & 0.949    &  & 0.471     & 0.939    & 1.000    & 1.000    \\
		50  &  & 0.049    & 0.054    & 0.041   & 0.036   &  & 0.236     & 0.700     & 0.962    & 1.000    &  & 0.840     & 0.999    & 1.000    & 1.000    \\
		100 &  & 0.050    & 0.051    & 0.049   & 0.034   &  & 0.498     & 0.957     & 1.000    & 1.000    &  & 0.991     & 1.000    & 1.000    & 1.000    \\
		200 &  & 0.038    & 0.036    & 0.031   & 0.035   &  & 0.828     & 0.999     & 1.000    & 1.000    &  & 1.000     & 1.000    & 1.000    & 1.000    \\ \bottomrule
	\end{tabular}
	\begin{tablenotes}\item \emph{Notes}: This table reports rejection frequencies for the $\mathrm{sup}W_{\bm{\delta}_0}$ test, which tests $H_0:\bm{\delta}_0=0$ against $H_1:\bm{\delta}_0 \in \{0.25,0.5\}$. The number of bootstrap replications is $B=300$. 											\end{tablenotes}
\end{table}

\begin{table}[h]
	\caption{Bias, RMSE and 95\% coverage results for $\hat\gamma$.}
	\hskip 12 pt
	\centering
	\footnotesize
	\par
	\begin{tabular}{@{}llllllllllllllll@{}}
		\toprule
		&  & \multicolumn{4}{l}{Bias}          &  & \multicolumn{4}{l}{RMSE}      &  & \multicolumn{4}{l}{Coverage}  \\ \cmidrule(lr){3-6} \cmidrule(lr){8-11} \cmidrule(l){13-16}
		$N$/$T$ &  & 10     & 25     & 50     & 100    &  & 10    & 25    & 50    & 100   &  & 10    & 25    & 50    & 100   \\ \midrule
		\multicolumn{16}{c}{$\alpha=0$}                                                                           \\
		25  &  & -0.161 & -0.052 & -0.014 & -0.006 &  & 0.280 & 0.059 & 0.011 & 0.002 &  & 0.917 & 0.925 & 0.905 & 0.801 \\
		50  &  & -0.073 & -0.019 & -0.004 & -0.001 &  & 0.092 & 0.016 & 0.002 & 0.000 &  & 0.891 & 0.891 & 0.800 & 0.546 \\
		100 &  & -0.025 & -0.004 & -0.001 & 0.000  &  & 0.025 & 0.001 & 0.000 & 0.000 &  & 0.880 & 0.761 & 0.546 & 0.316 \\
		200 &  & -0.006 & -0.002 & 0.000  & 0.000  &  & 0.005 & 0.001 & 0.000 & 0.000 &  & 0.788 & 0.539 & 0.310 & 0.146  \\
		\multicolumn{16}{c}{$\alpha=0.1$}                                                                                \\
		25  &  & -0.251 & -0.115 & -0.049 & -0.016 &  & 0.639 & 0.178 & 0.053 & 0.011 &  & 0.927 & 0.927 & 0.923 & 0.889 \\
		50  &  & -0.184 & -0.052 & -0.024 & -0.005 &  & 0.338 & 0.061 & 0.019 & 0.003 &  & 0.897 & 0.932 & 0.897 & 0.797 \\
		100 &  & -0.110 & -0.024 & -0.005 & -0.001 &  & 0.149 & 0.022 & 0.004 & 0.000 &  & 0.907 & 0.902 & 0.799 & 0.609 \\
		200 &  & -0.041 & -0.005 & -0.002 & -0.001 &  & 0.044 & 0.004 & 0.001 & 0.000 &  & 0.900 & 0.836 & 0.642 & 0.394  \\
		\multicolumn{16}{c}{$\alpha=0.2$}                                                                                \\
		25  &  & -0.348 & -0.214 & -0.107 & -0.046 &  & 1.347 & 0.457 & 0.163 & 0.048 &  & 0.930 & 0.942 & 0.937 & 0.913 \\
		50  &  & -0.310 & -0.135 & -0.066 & -0.020 &  & 0.924 & 0.238 & 0.088 & 0.018 &  & 0.905 & 0.932 & 0.934 & 0.906 \\
		100 &  & -0.236 & -0.087 & -0.031 & -0.012 &  & 0.553 & 0.119 & 0.030 & 0.008 &  & 0.917 & 0.934 & 0.917 & 0.860 \\
		200 &  & -0.158 & -0.052 & -0.017 & -0.004 &  & 0.287 & 0.058 & 0.014 & 0.002 &  & 0.909 & 0.911 & 0.890 & 0.739 \\ \bottomrule
	\end{tabular}
	
	\begin{tablenotes}	\item \emph{Notes}: The table reports bias and root mean squared error (RMSE) of $\hat\gamma$ and the empirical coverage of a 95\% confidence interval for $\gamma_0$. $\alpha$ refers to the rate of shrinking of the threshold effect $\bm{\delta}_0$.
	\end{tablenotes}
\end{table}

\addtolength{\tabcolsep}{-0.18em}
\begin{table}
	\caption{5\% size and power results for $t_{\theta_0}$.}
	\hskip 12 pt
	\centering
	\footnotesize
	\par
	\begin{tabular}{@{}llllllllllllllllllll@{}}
		\toprule
		& \multicolumn{4}{c}{Size of   $t_{\beta_{0,1}}$} &  & \multicolumn{4}{c}{Power of   $t_{\beta_{0,1}}$} &  & \multicolumn{4}{c}{Size of   $t_{\delta_{0}}$} &  & \multicolumn{4}{c}{Power of   $t_{\delta_{0}}$} \\ \cmidrule(lr){2-5} \cmidrule(lr){7-10} \cmidrule(lr){12-15} \cmidrule(l){17-20}
		$N$/$T$ & 10         & 25         & 50        & 100       &  & 10         & 25         & 50         & 100       &  & 10         & 25        & 50        & 100       &  & 10         & 25         & 50        & 100       \\ \midrule
		\multicolumn{20}{c}{$\alpha=0$}                                                                                                                                                                                      \\
		25  & 0.073      & 0.068      & 0.058     & 0.063     &  & 0.268      & 0.557      & 0.821      & 0.985     &  & 0.075      & 0.069     & 0.070     & 0.068     &  & 0.101      & 0.135      & 0.211     & 0.416     \\
		50  & 0.059      & 0.055      & 0.056     & 0.047     &  & 0.460      & 0.842      & 0.991      & 1.000     &  & 0.056      & 0.061     & 0.057     & 0.049     &  & 0.105      & 0.216      & 0.398     & 0.703     \\
		100 & 0.055      & 0.048      & 0.058     & 0.056     &  & 0.741      & 0.989      & 1.000      & 1.000     &  & 0.065      & 0.044     & 0.052     & 0.057     &  & 0.159      & 0.386      & 0.725     & 0.944     \\
		200 & 0.039      & 0.056      & 0.050     & 0.052     &  & 0.961      & 1.000      & 1.000      & 1.000     &  & 0.056      & 0.058     & 0.058     & 0.052     &  & 0.298      & 0.686      & 0.941     & 0.999     \\
		\multicolumn{20}{c}{$\alpha=0.1$}                                                                                                                                                                                    \\
		25  & 0.070      & 0.058      & 0.061     & 0.063     &  & 0.262      & 0.555      & 0.835      & 0.987     &  & 0.069      & 0.060     & 0.065     & 0.061     &  & 0.095      & 0.137      & 0.227     & 0.375     \\
		50  & 0.052      & 0.046      & 0.056     & 0.051     &  & 0.456      & 0.855      & 0.988      & 1.000     &  & 0.066      & 0.064     & 0.061     & 0.056     &  & 0.131      & 0.216      & 0.376     & 0.693     \\
		100 & 0.051      & 0.063      & 0.059     & 0.058     &  & 0.739      & 0.989      & 1.000      & 1.000     &  & 0.061      & 0.054     & 0.051     & 0.055     &  & 0.170      & 0.385      & 0.689     & 0.935     \\
		200 & 0.056      & 0.048      & 0.054     & 0.057     &  & 0.948      & 1.000      & 1.000      & 1.000     &  & 0.060      & 0.057     & 0.058     & 0.053     &  & 0.271      & 0.679      & 0.937     & 1.000     \\
		\multicolumn{20}{c}{$\alpha=0.2$}                                                                                                                                                                                    \\
		25  & 0.070      & 0.062      & 0.071     & 0.050     &  & 0.258      & 0.569      & 0.839      & 0.983     &  & 0.078      & 0.066     & 0.055     & 0.053     &  & 0.088      & 0.133      & 0.208     & 0.359     \\
		50  & 0.053      & 0.061      & 0.055     & 0.059     &  & 0.442      & 0.851      & 0.992      & 1.000     &  & 0.064      & 0.063     & 0.053     & 0.049     &  & 0.125      & 0.242      & 0.376     & 0.663     \\
		100 & 0.052      & 0.055      & 0.046     & 0.057     &  & 0.750      & 0.990      & 1.000      & 1.000     &  & 0.060      & 0.055     & 0.049     & 0.064     &  & 0.185      & 0.374      & 0.651     & 0.920     \\
		200 & 0.049      & 0.051      & 0.052     & 0.049     &  & 0.951      & 1.000      & 1.000      & 1.000     &  & 0.058      & 0.048     & 0.053     & 0.050     &  & 0.298      & 0.640      & 0.925     & 0.999     \\ \bottomrule
	\end{tabular}
	
	\begin{tablenotes}	\item \emph{Notes}: The table reports rejection frequencies of the $t_{\theta_0}$ test when testing $H_0: \beta_{0,1}=1$ against $H_1: \beta_{0,1}=0.9$ and $H_0: \bm{\delta}_0 =N^{-\alpha}$ against $H_1: \bm{\delta}_0 =N^{-\alpha}-0.1$. In the table, these tests are referred to as $t_{\beta_{0,1}}$ and $t_{\delta_{0}}$, respectively. 													\end{tablenotes}
\end{table}

\begin{table}[]
	\caption{Descriptive statistics.}
	\hskip 12 pt
	\centering
	\footnotesize
	\par
	\begin{tabular}{@{}llllllll@{}}
		\toprule
		Variable              & Obs   & Mean   & STD & Min     & Max       & CD  & Corr \\ \midrule
		\multicolumn{8}{c}{Full sample}\\
		GDP  & 3,922 & 1.918  & 4.504     & -41.587 & 60.091    & 72.87***  & 0.193  \\
		INF    & 3,922 & 1.462  & 1.384     & -12.686 & 9.372     & 127.03*** & 0.336  \\
		GOV      & 3,922 & 15.348 & 5.210     & 2.926   & 45.959    & 37.01***  & 0.098 \\
		TRA           & 3,922 & 76.571 & 59.455    & 2.699   & 442.620   & 86.47***  & 0.229 \\
		POP        & 3,922 & 1.644  & 1.263     & -16.881 & 16.626    & 71.7***  & 0.190  \\
		CAP  & 3,922 & 21.993 & 6.265     & 2.178   & 60.562    & 15.21***  & 0.040 \\
		\multicolumn{8}{c}{Developing countries}\\
		GDP    & 2,703 & 1.789  & 5.014     & -41.587 & 60.091    & 36.5***   & 0.140  \\
		INF    & 2,703 & 1.638  & 1.465     & -12.686 & 9.372     & 65.1***   & 0.323  \\
		GOV      & 2,703 & 13.854 & 4.974     & 2.926   & 45.959    & 10.11***   & 0.039\\
		TRA           & 2,703 & 72.604 & 50.288    & 2.699   & 442.620   & 35.47***  & 0.136 \\
		POP        & 2,703 & 2.075  & 1.223     & -16.881 & 16.626    & 95.77***  & 0.368 \\
		CAP  & 2,703 & 21.119 & 6.652     & 2.178   & 60.562    & 16.97***  & 0.065 \\
		\multicolumn{8}{c}{Industrialized countries}\\
		GDP    & 1,219 & 2.203  & 3.067     & -11.600 & 23.305    & 56.52***  & 0.488 \\
		INF    & 1,219 & 1.072  & 1.089     & -5.478  & 3.435     & 83.63***  & 0.411  \\
		GOV      & 1,219 & 18.659 & 4.072     & 8.171   & 27.935    & 40.55***  & 0.350 \\
		TRA           & 1,219 & 85.366 & 75.215    & 10.757  & 437.327   & 78.85***  & 0.681 \\
		POP        & 1,219 & 0.690  & 0.703     & -4.170  & 5.322     & 10.34***  & 0.089 \\
		CAP  & 1,219 & 23.929 & 4.769     & 10.411  & 54.274    & 32.94***  & 0.285 \\\bottomrule
	\end{tabular}
	\begin{tablenotes}\item \emph{Notes:} ``Obs'', ``Mean'', ``STD'', ``Min'' and ``Max'' refer to the total number of observations, sample average, the sample standard deviation, the sample minimum value and the sample maximum value, respectively. The column labelled ``CD'' reports the results obtained by applying Pesaran's (2021) test for cross-sectional correlation, while the column labelled ``Corr'' reports the average of all pairwise cross-correlations. The superscripts ``*'', ``**'' and ``***'' denote statistical significance at the 10\%, 5\% and 1\% levels, respectively. 	
	\end{tablenotes}
\end{table}

\begin{table}[]
	\caption{The estimated inflation--growth relationship.}
	\hskip 12 pt
	\centering
	\footnotesize
	\par
	\begin{tabular}{@{}lllll@{}}
		\toprule
		&  & Full sample & Developing         & Industrialized      \\ \midrule
		&  &                    &                    &                                \\
		Threshold (\%) &  & 3.476***           & 1.141***           & 3.401**           \\
		&  & {[}2.787, 4.336{]} & {[}1.029, 1.306{]} & {[}2.829, 6.619 {]}  \\
		&  &                    &                    &                       \\
		GOV (low inflation)      &  & -0.402***          & -0.416***          & -0.661***        \\
		&  & (0.081)            & (0.098)            & (0.190)             \\
		INF (low inflation) &  & 0.361**            & 0.391              & 0.185           \\
		&  & (0.159)            & (0.253)            & (0.381)            \\
		&  &                    &                    &                       \\
		GOV (high inflation)     &  & 0.132***           & 0.147***           & 0.098***            \\
		&  & (0.028)            & (0.040)            & (0.026)               \\
		INF (high inflation) &  & -1.702***          & -1.514***          & -1.810***          \\
		&  & (0.279)            & (0.347)            & (0.312)              \\
		&  &                    &                    &                             \\
		TRA        &  & 0.032***           & 0.025              & 0.022*                   \\
		&  & (0.013)            & (0.018)            & (0.013)              \\
		POP      &  & -0.752**           & -0.847***          & -1.104***         \\
		&  & (0.159)            & (0.137)            & (0.244)                     \\
		CAP        &  & 0.114***           & 0.117**            & -0.007             \\
		&  & (0.040)            & (0.050)            & (0.071)                \\
		&  &                    &                    &                           \\
		$\mathrm{sup}W_{\delta_0}$       &  & 37.109***          & 23.763**           & 13.257            \\
		Rank classifier     &  & Pass                & Pass                & Pass                   \\
		Residual CD test   &  & -1.059             & -1.0105            & -1.840*               \\
		Observations             &  & 3922               & 2,703              & 1,219                  \\ \bottomrule
	\end{tabular}
	\begin{tablenotes} \item \emph{Notes}: The coefficients of INF and GOV are allowed to break depending on whether the threshold variable (INF) is above or below the estimated threshold. $\mathrm{sup}W_{\delta_0}$ tests the null hypothesis of no threshold effects. The row labelled ``Rank classifier'' states if Assumption \ref{ass:load} passes or fails according to the rank condition classifier of De Vos et al. (2024). The row labelled ``Residual CD test'' report the results obtained by applying Juodis and Reese's (2022) version of the CD test, which tests for remaining cross-sectional correlation in the regression residuals. The numbers reported within round brackets are the estimated standard errors, while those reported within square brackets are the estimated 95\% confidence intervals. The superscripts ``*'', ``**'' and ``***'' denote statistical significance at the 10\%, 5\% and 1\% levels, respectively.									\end{tablenotes}
\end{table}
\end{document}